\documentclass{pasj00}

\begin{document}
\SetRunningHead{M. Sasada et al.}{Photopolarimetric observations of AO~0235$+$164 and PKS~1510$-$089}
\Received{2010/11/16}
\Accepted{2011/02/04}

\title{Prominent Polarized Flares of the Blazars AO~0235$+$164 and PKS~1510$-$089}

\author{
Mahito \textsc{Sasada}\altaffilmark{1},
Makoto \textsc{Uemura}\altaffilmark{2},
Yasushi \textsc{Fukazawa}\altaffilmark{1},
Koji S. \textsc{Kawabata}\altaffilmark{2},\\
Yuki \textsc{Ikejiri}\altaffilmark{1},
Ryosuke \textsc{Itoh}\altaffilmark{1},
Masayuki \textsc{Yamanaka}\altaffilmark{2},
Kiyoshi \textsc{Sakimoto}\altaffilmark{1},
Takashi \textsc{Ohsugi}\altaffilmark{2},\\
Michitoshi \textsc{Yoshida}\altaffilmark{2},
Shuji \textsc{Sato}\altaffilmark{3},
and Masaru \textsc{Kino}\altaffilmark{3}}

\altaffiltext{1}{Department of Physical Science, Hiroshima University, Kagamiyama 1-3-1, Higashi-Hiroshima 739-8526}
\email{sasada@hep01.hepl.hiroshima-u.ac.jp}
\altaffiltext{2}{Astrophysical Science Center, Hiroshima
University, Kagamiyama 1-3-1, Higashi-Hiroshima 739-8526}
\altaffiltext{3}{Department of Physics, Nagoya University, Furo-cho, Chikusa-ku, Nagoya 464-8602}


%

\KeyWords{BL Lacertae Objects: individual: AO~0235$+$164, PKS~1510$-$089
--- polarization --- infrared: general} 

\maketitle

\begin{abstract}
We report on multi-band photometric and polarimetric observations of the
 blazars AO~0235$+$164 and PKS~1510$-$089. These two blazars were active
 in 2008 and 2009, respectively. In these active states, prominent short
 flares were observed in both objects, having amplitudes of $>1$~mag
 within 10~d. The $V-J$ color became bluer when the objects were
 brighter in these flares. On the other hand, the color of PKS~1510$-$089 
 exhibited a trend that it became redder when it was brighter, except for 
 its prominent flare. This redder-when-brighter trend can be explained by 
 the strong contribution of thermal emission from an accretion disk. 
 The polarization degree increased at the flares, and 
 reached $>25$~\% at the maxima. We compare these flares in AO~0235$+$164 
 and PKS~1510$-$089 with other short flares which were detected by our 
 monitoring of 41 blazars. Those two flares had one of the largest 
 variation amplitudes in both flux and polarization degree. 
 Furthermore, we found a significant positive correlation between the 
 amplitudes of the flux and polarization degree in the short flares.
 It indicates that the short flares originate from the region where 
 the magnetic field is aligned.
\end{abstract}

\section{Introduction}
Blazars emit nonthermal radiation over a wide frequency range from radio
to gamma-ray bands. There are two components in the emission from
blazars \citep{Fossati98}. The low energy component from radio to
optical or X-ray bands is attributed to synchrotron radiation emitted by
relativistic electrons in jets. The high energy component is likely to
be inverse-Compton scattering radiation. The source of its seed photons
has not been identified yet: they may come from the synchrotron
radiation (e.g. \cite{Jones74}; \cite{Marscher85}) or have external
origin (e.g. \cite{Dermer93}; \cite{Sikora94}).

The synchrotron emission from blazars is occasionally highly polarized
in the optical band \citep{Angel80}. Since the polarization gives us a
clue to probe magnetic field of a nonthermal radiation source,
observations of temporal variations of the polarization are important
for study of the structure of a blazar jet. However, the temporal
behavior of the polarization is known to be complex. The polarization
varies in an erratic manner in most blazars (e.g. \cite{Angel80}). On
the other hand, systematic variations have also been reported in several
blazars, for example, rotations of the polarization (e.g. \cite{Qian91};
\cite{Sillanpaa93}). The origin of the rotations of polarization is
still an open question; \citet{Marscher08} have suggested a helical
structure of the magnetic field based on a rotation observed in
BL~Lac. \citet{Abdo10a} have proposed that a rotation of polarization in
3C~279 indicates a bending of the jet. \citet{Jorstad07} have reported
positive correlations between the total flux, fractional polarization,
and polarization position angle in the radio and optical bands in 15
active galactic nuclei. 

It has been suggested that the erratic behavior of polarization is due 
to the composition of several polarization components \citep{Moore82}. 
From the data of OJ~287 from 2005 to 2009, \citet{Villforth10} have 
reported that its observed polarization behavior could be explained by 
separating the jet emission into two components: an optical 
polarization core and chaotic jet emission. \citet{Uemura10} have 
separated a long-term trend from observed temporal variations of 
polarization in blazars using a Bayesian approach.

AO~0235$+$164 is one of the most famous blazars, showing violent
variability whose amplitude is larger than 1~mag
(e.g. \cite{Rieke76}). Its spectrum shows two absorption-line systems at
$z=0.94$ and $z=0.524$. The former has been considered as that of
AO~0235$+$164, and the latter is probably attributed to a foreground
galaxy (\cite{Roberts76}; \cite{Burbidge76}). The object experienced
large amplitude outbursts in past. Radio outbursts repeated
quasi-regularly with a periodicity of $\sim$5.7 years
\citep{Raiteri01}. The object is highly polarized in the optical band,
and the polarization parameters also vary violently. The highest
polarization degree has been reported to be 43.9~\% \citep{Impey82}.

PKS~1510$-$089 is a radio-loud, highly polarized quasar at
$z=0.361$. The object is detectable from the radio to gamma-ray bands
(e.g. \cite{Kataoka08}; \cite{Abdo10b}). The spectral energy
distribution of the object has a bump structure over a synchrotron
component in the ultraviolet (UV) band (\cite{Singh97};
\cite{Kataoka08}). The UV bump is thought to be thermal emission from an
accretion disk. Polarization parameters of the object have also been
varied. \citet{Marscher10} reported that the polarization rotated
consecutively for 50$\pm$10~d.

In the above two blazars, we detected prominent short-term flares with a
timescale of $\sim 10$~d, and found that they were associated with
violent variations in polarization. In this paper, we report their light
curves, color and polarization variations, and discuss the
characteristics in polarization variations of such short-term
flares. This paper is arranged as follows: In section~2, we present our
observation method and analysis. In section~3, we report the result of
the photometric and polarimetric observations of these objects. In
section~4, we first describe long-term and short-term variation
components indicated by the flux and color variations. Second, we
discuss a correlation between the amplitudes of the flux and
polarization degree in short-term flares in 41 blazars. The conclusion
is drawn in section~5.

\section{Observation}
We performed monitoring of AO~0235$+$164 and PKS~1510$-$089 in the
2008--2009 seasons using TRISPEC attached to the Kanata 1.5-m telescope
at Higashi-Hiroshima Observatory. TRISPEC (Triple Range Imager and
SPECtrograph) has a CCD and two InSb arrays, which enable
photopolarimetric observations in an optical and two NIR bands
simultaneously \citep{Watanabe05}. Unfortunately, the $K_{\rm S}$-band
readout system of TRISPEC was out of action. Thus, we obtained only 
$V$- and $J$-band data. A unit of the observing sequence 
consisted of successive exposures at four position angles of a 
half-wave plate; 0\arcdeg, 45\arcdeg, 22.\arcdeg5 and 67.\arcdeg5. A 
set of polarization parameters was derived from each set of the four 
exposures. 

The integration time for an exposure depends on the sky condition and
the brightness of the objects. Typical integration times were 120 and
108~sec in AO~0235$+$164, and 150 and 135~sec in PKS~1510$-$089 in the
$V$ and $J$ bands, respectively. 

All images were bias-subtracted and flat-fielded, and we performed an
aperture photometry with a Java-based package. We performed a
differential photometry with comparison stars taken in the same frames
of AO~0235$+$164 and PKS~1510$-$089. The positions of the comparison 
stars are
R.A.=$\timeform{02h38m32s.31}$, Dec.=$\timeform{+16D35'59''.7}$ and
R.A.=$\timeform{15h12m53s.19}$, Dec.=$\timeform{-09D03'43''.6}$
(J2000.0), respectively. Their magnitudes are $V$=12.720, 13.282 and
$J$=11.221, 12.205, respectively (\cite{Gonzalez-perez01};
\cite{Skrutskie06}). The constancy of the comparison stars was checked
by another neighbor stars in the same frames. The position of the check
stars are R.A.=$\timeform{02h38m36s.70}$,
Dec.=$\timeform{+16D36'26''.6}$ and R.A.=$\timeform{15h12m51s.64}$,
Dec.=$\timeform{-09D05'24.''1}$ (J2000.0), respectively. No significant flux
variation was observed in the comparison stars over 0.16 and 0.07 mag
in the $V$ band during our observation period. 

We confirmed that the instrumental polarization was smaller than 0.1~\%
in the $V$ band using observations of unpolarized standard stars. We,
thus, applied no correction for it. The zero point of the polarization
angle is corrected for in the equatorial coordinate system (measured 
from north to east) by observing the polarized stars, HD~19820 and 
HD~25443 \citep{Wolff96}.

\section{Result}
\subsection{AO~0235$+$164}
\begin{figure*}
  \begin{center}
 \FigureFile(160mm,100mm){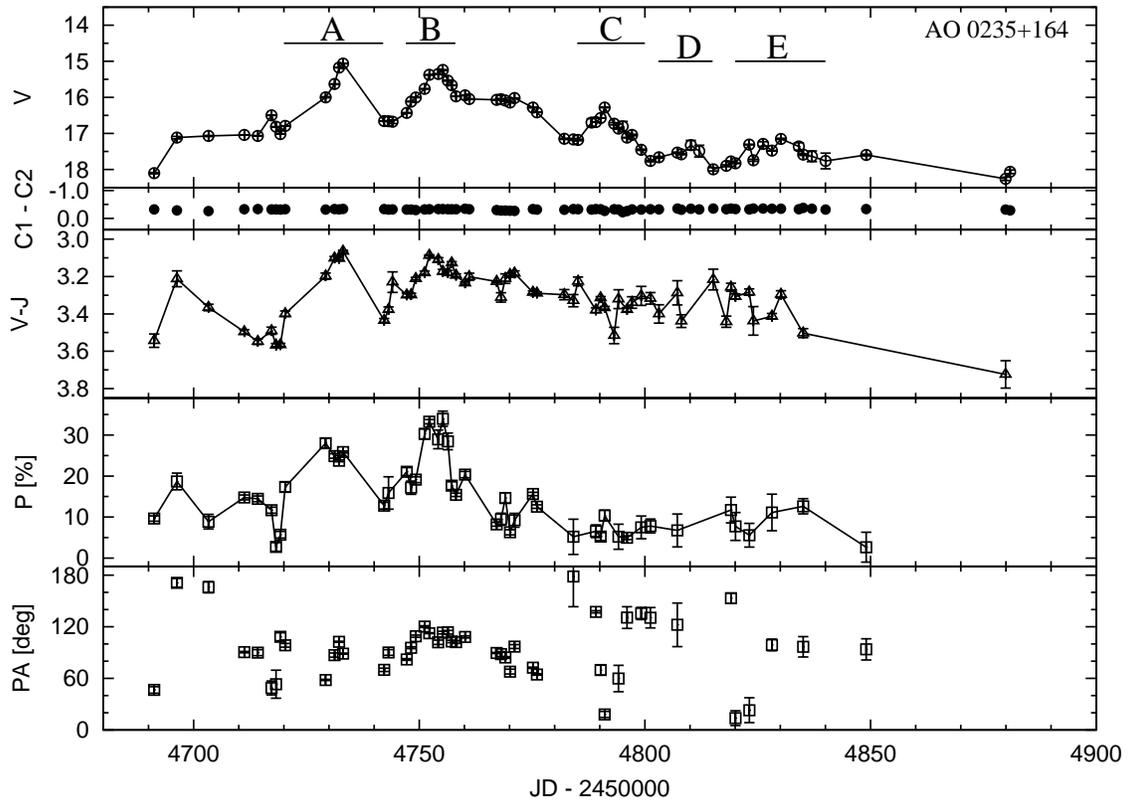}
  \end{center}
  \caption{Light curve of AO~0235$+$164 in the $V$ band, color 
  variations of $V-J$, temporal variations of the polarization 
  degree (\%) and angle (degree). We also show the relative 
  magnitude, C1-C2, between the comparison star (C1) and the 
  check star (C2).}%
  \label{fig1}
\end{figure*}

We had performed the photopolarimetric monitoring of AO~0235$+$164 from
Aug. 12, 2008 to Feb. 18, 2009. Figure~1 shows the light curve of the
object in the $V$ band, $V-J$ color variation, and temporal
variations of the polarization parameters in the $V$ band. We also show
relative magnitude between the comparison and the check stars. In
the top panel of figure~1, the object showed the violent
variability. The object had been active for 153~d from JD~2454696 to
2454849, in which it was brighter than $V=18.0$. The peak flux of the
object was $V$=15.067$\pm$0.004 (JD~2454733). The faintest state of the
object has been reported at $V\sim20$ for 32 years from 1975 to 2007
(\cite{Raiteri01}; \cite{Raiteri05}; \cite{Raiteri06};
\cite{Raiteri08}). Thus, the amplitude of the flux variation was about
5~mag. The light curve shows several short flares during the active
state. We can see five flares labelled in figure~1 as ``A'' (from
JD~2454720 to 2454742), ``B'' (from JD~2454747 to 2454758), ``C'' (from
JD~2454785 to 2454801), ``D'' (from JD~2454803 to 2454815) and ``E''
(from JD~2454820 to 2454840).  

The $V-J$ color was variable in our monitoring period,
$\Delta\;(V-J)\sim\;0.5$. In flares ``A'' and ``B'', the color became
bluer. The object, thus, showed a bluer-when-brighter trend in these
prominent flares. The color also became bluer in JD~2454696, at the
onset of the active state when the object was brightened rapidly. On the 
other hand, no clear bluer-when-brighter trend was associated with 
flares ``C'', ``D'' and ``E''.In the faintest state in JD~2454879, the 
object was the reddest. 

The polarization degree was variable, and distributed from 0~\% to
$\sim$30~\%. In JD~2454755, the polarization degree was the highest,
34$\pm$2~\%. The polarization degree was correlated with the flux in the
flares ``A'' and ``B''. The peaks of the polarization degree in the flares
``A'' and ``B'' were higher than 25~\%. However, the polarization degrees
were not high during the flares ``C'', ``D'' and ``E'', less than 16~\%. 

\begin{figure}
  \begin{center}
 \FigureFile(80mm,100mm){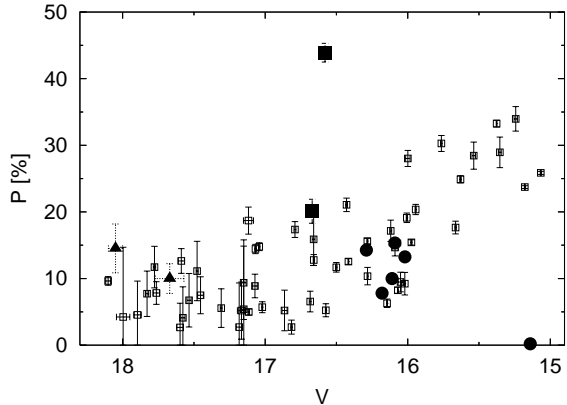}
  \end{center}
  \caption{Correlation between the magnitude and the polarization degree 
  in the $V$ band. Open squares, filled squares, triangles and 
  circles denote the observations with the Kanata telescope, reported 
  by \citet{Impey82}, \citet{Mead90} and \citet{Takalo92}, 
  respectively.}%
  \label{fig2}
\end{figure}

Figure~2 shows the correlation between the flux and the polarization
degree in the $V$ band. We show our observations by open squares,
and the data reported by \citet{Impey82}, \citet{Mead90} and
\citet{Takalo92} by filled squares, triangles and circles,
respectively. The $V$-band flux was moderately correlated with the
polarization degree when the object was brighter than $V=17$. The flux
was uncorrelated with the polarization degree when it was fainter than
$V=17$, while the error of the polarization degree is large in the faint
state. A correlation coefficient between the flux and polarization
degree was calculated to be $0.74^{+0.09}_{-0.14}$ (95~\% confidence 
interval) in
the whole data. In the case of the data brighter than 17~mag., the
correlation coefficient was $0.76^{+0.11}_{-0.20}$.

In figure~2, most of the data reported in previous studies were
consistent with our data, while there are two exceptions. First,
\citet{Impey82} reported that AO~0235$+$164 underwent a polarization
burst, reaching a polarization degree of 43.9$\pm$1.4~\% when it was
$V=16.58$. In our data, the object exhibited polarization degrees lower
than 19~\% around $V=16.5$. Second, \citet{Takalo92} reported a low
polarization degree of 0.21~\% when the object was in a bright state at
$V=15.14$, while our data show high polarization degrees in such a
bright state. We discuss the behavior of polarization degree in 
section~4.2.

\begin{figure}
  \begin{center}
 \FigureFile(80mm,100mm){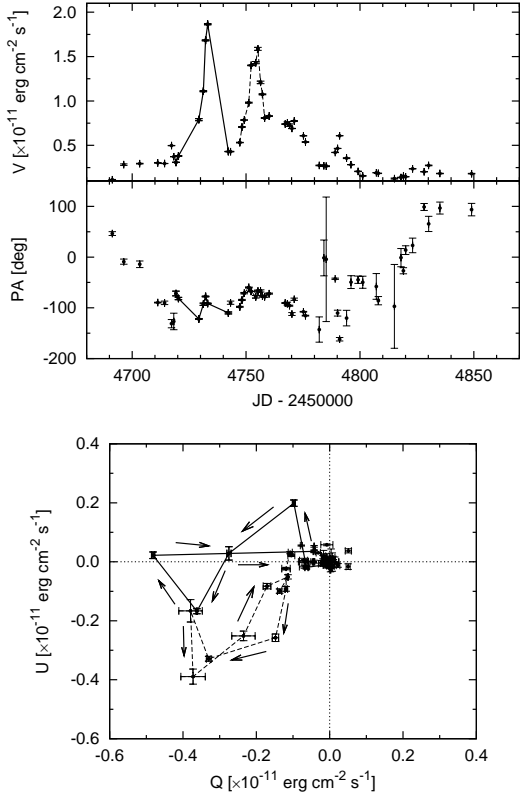}
  \end{center}
  \caption{Top panel: Light curve of AO~0235$+$164 in the $V$ band. 
  The flux is shown in $\nu\;F\nu$ and the unit of the flux is 
  ${\rm erg\;cm^{-2}\;s^{-1}}$. The flux was calculated with 
  1.98$\times\;10^{-5}\;{\rm\;erg\;cm^{-2}\;s^{-1}}$ 
  at 0~mag in the $V$ band. The solid and dashed lines represent the 
  flares ``A'' (from JD~2454719 to 2454743) and ``B'' (from 
  JD~2454747 to 2454761). Middle panel: The temporal variation of the 
  polarization angle in the $V$ band. For details of the correction 
  of the angle, see the text. Bottom panel: The $QU$ plane in the $V$ 
  band. Units of $Q$ and $U$ are ${\rm\;erg\;cm^{-2}\;s^{-1}}$.}%
  \label{fig3}
\end{figure}

Figure~3 shows the light curve (top panel), the temporal variation of
the corrected polarization angle (middle panel) and the temporal
variations of the polarization in the $QU$ plane (bottom panel) in the
$V$ band. We calculated the flux, $\nu\;F\nu$, with
1.98$\times\;10^{-5}\;{\rm\;erg\;cm^{-2}\;s^{-1}}$ at 0~mag in the $V$
band \citep{Fukugita95}. In figure~3, solid and dashed lines indicate
the flares ``A'' and ``B'', respectively. The polarization angle shown
in the middle panel was corrected to enable the angle to have values
over a range of 0---180$^{\circ}$. In general, the observed polarization
angle is defined from 0$^{\circ}$ to 180$^{\circ}$. We corrected the
polarization angle assuming that the difference between the polarization
angles of the temporal adjacent data should be less than
90$^{\circ}$. We defined this difference as 
$|\Delta\;PA_{n}|=|PA_{n+1}-PA_{n}|-\sqrt{\delta\;{PA_{n+1}}^{2}+\delta\;{PA_{n}}^{2}}$, 
where $PA_{n+1}$ and $PA_{n}$ were the n+1- and n-th polarization angles
and $\delta\;PA_{n+1}$ and $\delta\;PA_{n}$ were their errors. If
$|\Delta\;PA_{n}|$ is smaller than 90$^{\circ}$, no correction was
performed. If $\Delta\;PA_{n}$ is smaller than $-90^{\circ}$, we add
180$^{\circ}$ to $PA_{n+1}$. If $\Delta\;PA_{n}$ is larger than
90$^{\circ}$, we add $-180^{\circ}$ to $PA_{n+1}$. As shown in the
middle panel, the polarization angles both in the flares ``A'' and ``B''
were $-$122---$-$72 and $-$98---$-$60. The polarization angles of these
flares were, hence, different from each other. The polarized flux increased
with the flares ``A'' and ``B'', respectively, as can be seen in the
bottom panel. Thus, these flares had specific polarization components,
which were developed and diminished through the flares. The corrected
polarization angle from JD~2454800 to 2454860 seems to rotate
gradually. However, it is possible that this rotation was spurious
event, because the polarization degree was low and the object was faint
during the rotation. The error of the polarization angle was high when
the polarization degree was low. We need observations with a higher
sampling and smaller errors in order to confirm a long-term rotation
event like this.

\subsection{PKS~1510$-$089}
\begin{figure*}
  \begin{center}
 \FigureFile(160mm,100mm){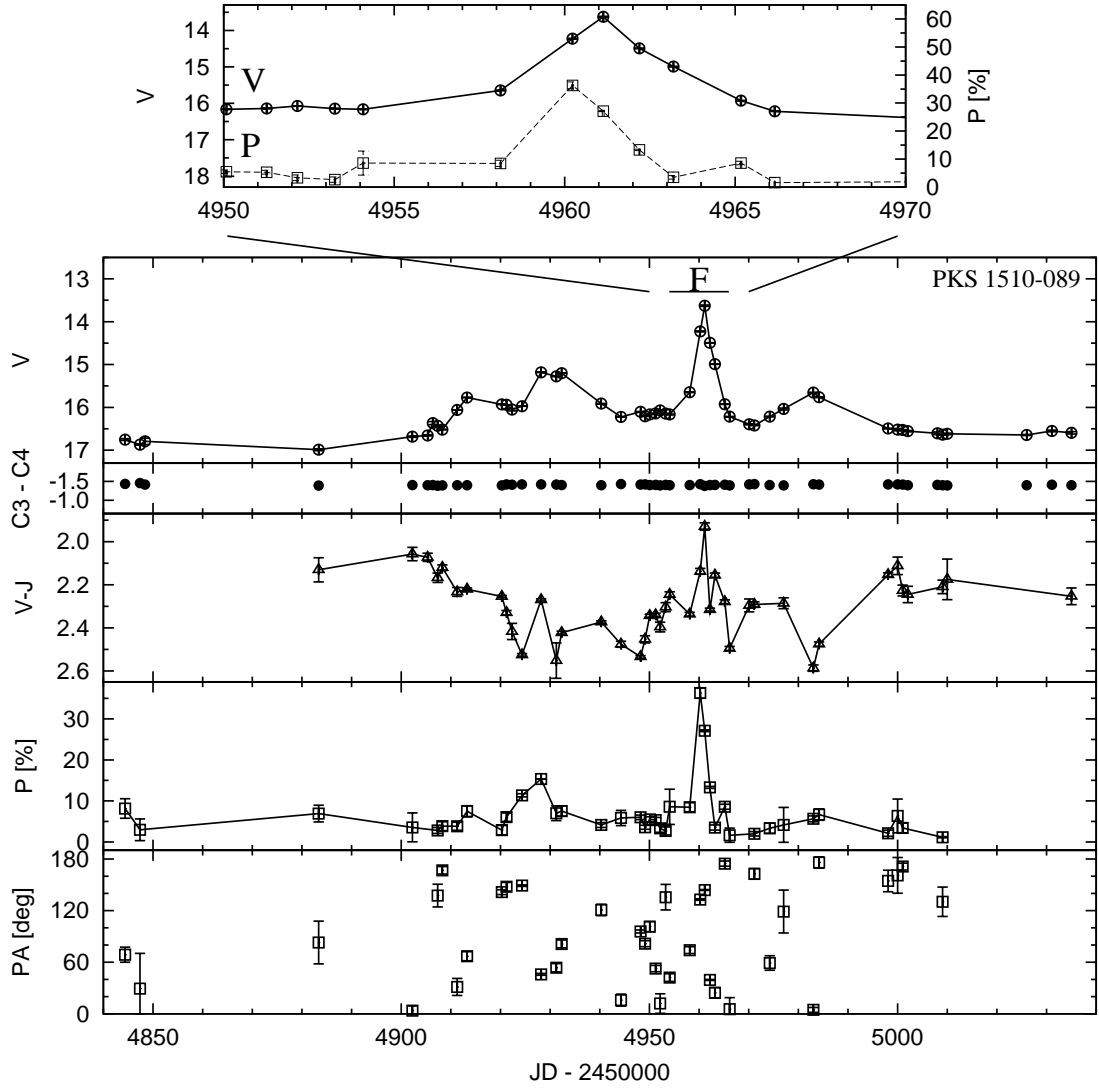}
  \end{center}
  \caption{Light curve of PKS~1510$-$089 in the $V$ band, color 
  variations of $V-J$, temporal variations of the polarization 
  degree (\%) and angle (degree). The top panel shows the enlarged 
  light curve and polarization degree around flare ``F''. We also 
  show the relative magnitude, C3-C4, between the comparison star 
  (C3) and the check star (C4).}%
  \label{fig4}
\end{figure*}

We had performed the photopolarimetric monitoring of PKS~1510$-$089 from
Jan. 12 to Jul. 22 in 2009. Figure~4 shows the light curve of
PKS~1510$-$089 in the $V$ band, $V-J$ color variation, and temporal
variations of the polarization parameters in the $V$ band. The object
has been brightened since JD~2454900, and a peak of the flux was once
recorded in JD~2454928. After that, a prominent flare occurred around
JD~2454960. We labelled this flare as ``F''  (from JD~2454954 to
2454966). In this flare, the peak of the flux reached about $V$=13.6~mag
in JD~2454961, and the amplitude of the magnitude between the peak and
the bottom of the flare in JD~2454954 was $\Delta\;V$=2.5~mag. After the
flare peak, the flux was fallen down to $V$=16.2~mag in
JD~2454966. Thus, the flux varied by ten times within 10~d. The flux
again rose after this flare, and the flux reached another peak at about
$V$=15.7~mag in JD~2454966. After this peak, the flux faded gradually. 

The $V-J$ color was about 2.1 in a faint state in JD~2454900. When the
object became brighter, the $V-J$ color got redder to 2.5. This
``redder-when-brighter'' trend was saturated at about $V-J$=2.5. On the
other hand, the color during the flare ``F'' showed a
bluer-when-brighter trend. The $V-J$ color reached 1.9 in this flare. 

Polarization parameters also varied in the whole observing period. The
polarization angle was distributed from 0 to 180$^{\circ}$. The
polarization degree varied less than 10~\% in a faint state from
JD~2454844 to 2454911. The polarization degree increased with the flux
from JD~2454920, and reached 15~\% at the maximum of a small flare in
JD~2454928. In the rising phase of the flare ``F'', the polarization
degree rapidly increased, and reached its peak about 36~\% in
JD~2454960, in which the flux was still in a rising phase (see the top
panel of figure~4). Thus, the peak of the polarization degree was 1~day
earlier than that of the flux. In the decay phase of the flux, the
polarization degree had already been low, about 13~\% in
JD~2454962. After the flare, the polarization degree did not vary more
than 10~\%. 

\begin{figure}
  \begin{center}
 \FigureFile(80mm,100mm){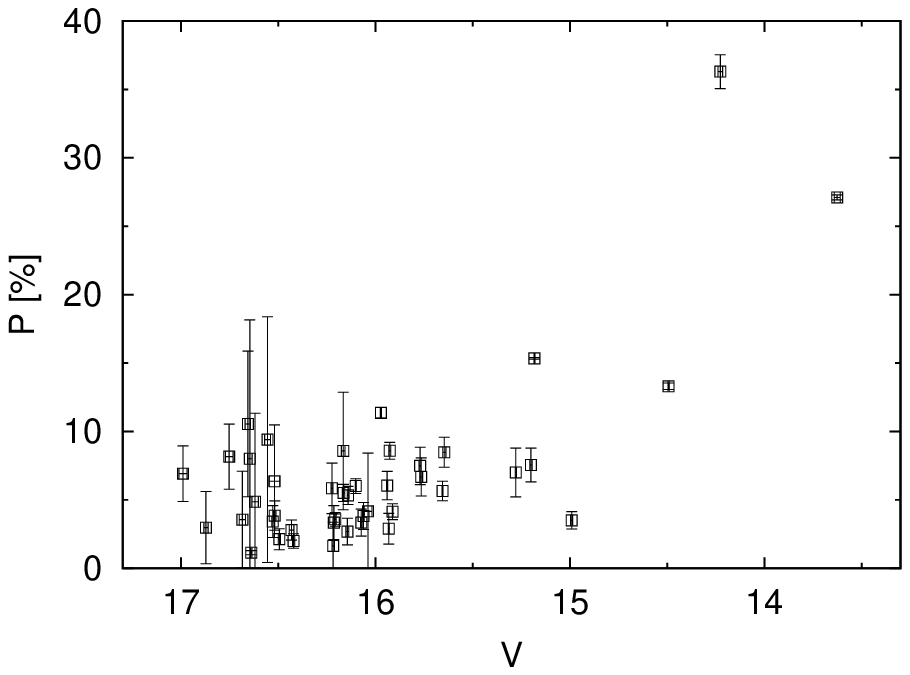}
  \end{center}
  \caption{Correlation between the magnitude and the polarization degree 
  in the $V$ band in PKS~1510$-$089.}%
  \label{fig5}
\end{figure}

Figure~5 shows a correlation between the flux and polarization degree
in the $V$ band in PKS~1510$-$089. The correlation is quite similar to
that observed in AO~0235$+$164; a positive correlation was clearly seen 
in a bright state, and in addition, no high polarization was detected 
when the object were faint. The correlation coefficient was calculated 
to be 0.70$^{+0.12}_{-0.19}$ using the whole data set in PKS~1510$-$089. 
The correlation coefficient using the data brighter than 15.5~mag was 
0.77$^{+0.19}_{-0.73}$ 

\begin{figure}
  \begin{center}
 \FigureFile(80mm,100mm){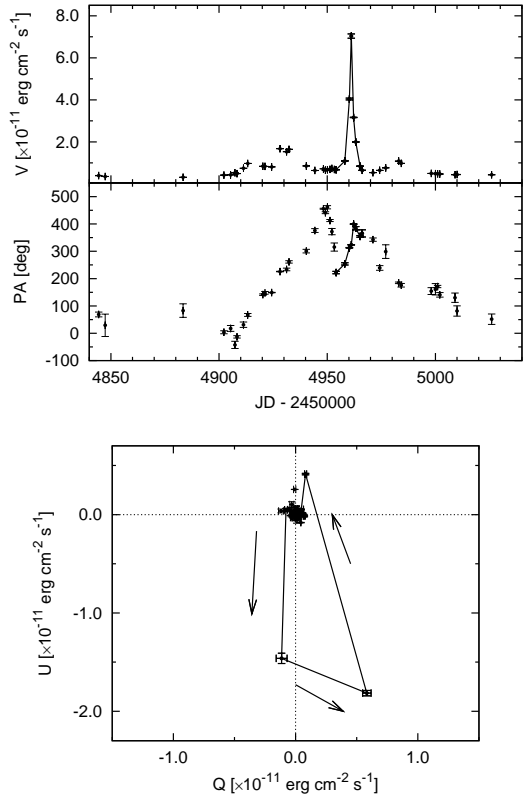}
  \end{center}
  \caption{Top panel: Light curve of PKS~1510$-$089 in the $V$
 band. The flux is shown in $\nu\;F\nu$ and the unit of the 
 flux is ${\rm erg\;cm^{-2}\;s^{-1}}$. The solid line
 represents the period of the flare ``F''. Middle panel: Corrected
 temporal variation of the polarization angle. For details of the
 correction, see the text of section~3.2. Bottom panel: The $QU$ plane
 in the $V$ band. Units of the $Q$ and $U$ are also ${\rm
 erg\;cm^{-2}\;s^{-1}}$. }%
  \label{fig6}
\end{figure}

Figure~6 shows the light curve (top panel), the corrected polarization
angle (middle panel) and the $QU$ plane (bottom panel) of PKS~1510$-$089
in the $V$ band. The solid line of figure~6 represents the period of the
flare ``F''. We applied the same correction as described in section~3.2
for the polarization angle. In the middle panel of figure~6, we show
corrected polarization angle. As can be seen in this panel, the
polarization angle rotated in a positive direction from $-40$ to
461$^{\circ}$ from JD~2454907 to 2454950. After a short rotation in a
negative direction, it again showed a positive rotation by 173$^{\circ}$
during flare ``F''. The $QU$ plane in the bottom panel of figure~6 also
showed the rotation of the polarization during the flare. The
polarization parameters, $Q$ and $U$, were variable dramatically during
the flare. \citet{Marscher10} reported that the polarization angle of
this object had rotated for 50~d from  JD~2454910. Our observation
confirmed this event. The behaviors of polarization angle were different
between our data and that of \citet{Marscher10} during
JD~2454951--2454960. This inconsistency is possibly caused by a
difference in correction method for the polarization angle or in data
sampling rate during this period of time. Therefore, we should be careful 
in treating the successive rotation trend of the polarization angle.

\section{Discussion}
\subsection{Short- and Long-Term Variation Components}
During the flares ``A'' and ``B'' of AO~0235$+$164, the $V-J$ color
became bluer. On the other hand, the color did not become bluer clearly
during the flares ``C'', ``D'' and ``E''. The lack of the color change
in the later flares might be partly due to a low signal-to-noise ratio
of the photometric data. It is also possible that there was another
variation component having a color and a timescale different from those
of the flare component. The object was $V\sim\;17.0$ just before the
onset of the flare ``A'' ($\sim$ JD~2454719). It was fainter than
$V=17.0$ after the flare ``C''. If the color follows the
bluer-when-brighter trend, the color in the early phase just before the
flare ``A'' should be bluer than that in the late phase after the flare
``C''. However, the color in the early phase, $V-J$=3.5---3.6, was
redder than that in the late phase, $V-J$=3.3---3.5. This behavior may
indicate that there is a variable component which was red in the early
phase, then changed to blue in the late phase. The bluer-when-brighter 
trend was not seen throughout the active state. The light curve suggests
that the short-term flares were superimposed on an underlying long-term
variation component. The observed color variations can be explained by
this two-component scenario; the observed color was a composition of the
bluer-when-brighter short-term flares and long-term component 
having the color variations as mentioned above. \citet{Raiteri06} have 
reported that AO~0235$+$164 possibly showed a gradually reddening trend 
from 2002 to 2005 although the mean flux were similar in those years. 
This may also support the presence of the color-variation component on 
a long timescale.

We tried to separate these two components and to improve the
color--magnitude correlation in the short-term flares. We constructed
the long-term component using the linear interpolation of the apparent 
minima of short-term flares in the light curve. Panel~(a) of figure~7 
shows the light curves of the observed data, assumed long-term component 
and the residual short-term component. Panel~(b) shows the ratio of the
flux between the $V$ and $J$ bands ($\nu\;F_{\nu\;V}/\nu\;F_{\nu\;J}$) of
them. We assumed a long-term component which becomes gradually bluer
with time, as shown in the figure. Panel~(c) and (d) of figure~8 show the
flux--color diagrams between $\nu\;F_{\nu\;V}$ and 
$\nu\;F_{\nu\;V}/\nu\;F_{\nu\;J}$. The correlation coefficient between 
log$(\nu\;F_{\nu\;V})$ and $\nu\;F_{\nu\;V}/\nu\;F_{\nu\;J}$
was calculated to be 0.79$^{+0.08}_{-0.11}$ for the separated short-term
component, which was higher than than the observed one,
0.68$^{+0.11}_{-0.15}$. Thus, it demonstrates that the observed color
behavior can be explained with the two components, namely the
bluer-when-brighter short-term flares and the long-term component.

\begin{figure}
  \begin{center}
 \FigureFile(80mm,100mm){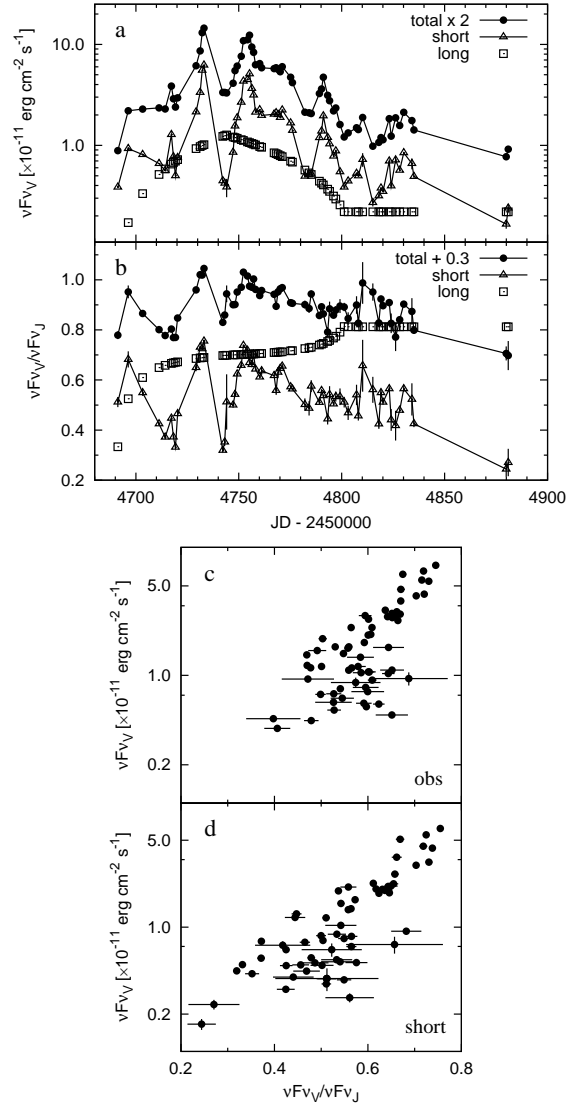}
  \end{center}
  \caption{Light curves, variations of flux ratio and flux--color
 diagrams in AO~0235$+$164. Panel~(a) shows the light curves of the
 observed data, the short-term and the long-term components,
 respectively. The flux is multiplied by a factor for readability. 
 Panel~(b) shows the variations of $\nu\;F_{\nu\;V}/\nu\;F_{\nu\;J}$ 
 of the observed data, the short-term and the long-term components. 
 We add 0.3 to the observed data of $\nu\;F_{\nu\;V}/\nu\;F_{\nu\;J}$
 for readability. The filled circle, triangle and open 
 square in panel~(a) and (b) show the observed data, the short-term 
 and long-term components, respectively. Panel~(c) and (d) show the
 flux--color diagrams in the observed data and the short-term
 component.}%
 \label{fig7} 
\end{figure}

The $V-J$ color became bluer during the flare ``F'' of
PKS~1510$-$089. On the other hand, the color during the active state
from JD~2454905 to 2454998 exhibited a redder-when-brighter trend. The
redder-when-brighter trend has been reported in a few blazars, for
example, 3C~454.3 (e.g. \cite{Miller81}). As mentioned before,
PKS~1510$-$089 has a bump structure in the spectral energy distribution
in the UV region (\cite{Singh97}; \cite{Kataoka08}). The
redder-when-brighter trend indicates that a synchrotron component, which
was redder than the UV bump component, became dominant during the active
state. The bluer-when-brighter trend in the flare ``F'' indicates that
the flare component was bluer than the underlying synchrotron
component. Thus, we can interpret the observed color behavior as a
composition of two synchrotron components; one was the long-term
component having a relatively red color, and the other was the
short-term flare component having a blue color. 

We also tried to separate the short-term flares and long-term component from
the observed data of PKS~1510$-$089. First, we estimated the $V$- and
$J$-band fluxes of the UV bump component using the inter- and
extrapolation of the data from $10^{14.45}$ to $10^{14.6}$~Hz (from 
7500 to 10600~\AA) reported
by \citet{Neugebauer79}. We subtracted the UV bump component from the
observed data, assuming the component was constant over time. After the
subtraction of the UV bump component, we separated the short-term and
long-term components in a similar way to the case of
AO~0235$+$164. Panel~(a) and (b) of figure~8 show the light curve in the
$V$ band and $\nu\;F_{\nu\;V}/\nu\;F_{\nu\;J}$ of the subtracted data,
the short-term and long-term components. The 
$\nu\;F_{\nu\;V}/\nu\;F_{\nu\;J}$ of the assumed long-term
component becomes bluer in its decay phase (JD~2454920---2455000), as that 
in AO~0235$+$164. Panel~(c), (d) and (e) show the diagrams between
$\nu\;F_{\nu\;V}$ and $\nu\;F_{\nu\;V}/\nu\;F_{\nu\;J}$ of the observed,
UV-bump subtracted data and short-term component. The correlation 
coefficients between log$(\nu\;F_{\nu\;V})$ and
$\nu\;F_{\nu\;V}/\nu\;F_{\nu\;J}$ are 0.00$\pm 0.30$,
0.68$^{+0.15}_{-0.10}$ and 0.88$^{+0.05}_{-0.08}$ for the observed, UV-bump
subtracted data, and short-term component, respectively. The correlation
is improved by assuming the long-term component and separating the
short-term flare component from the observed data.

\begin{figure}
  \begin{center}
 \FigureFile(80mm,100mm){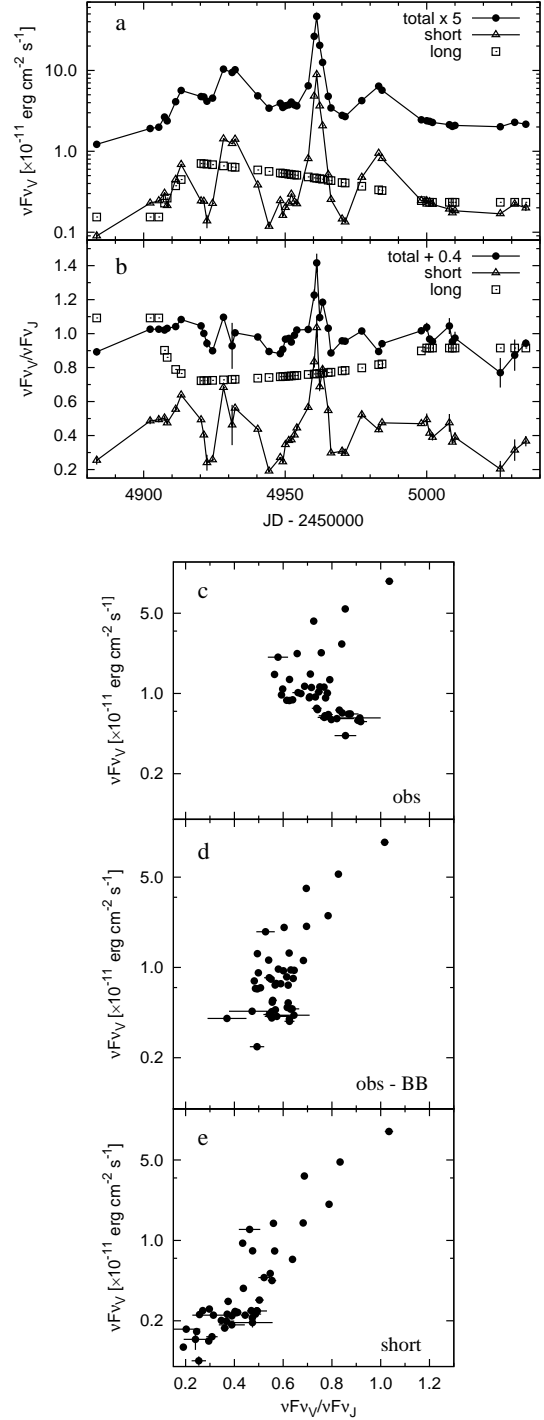}
  \end{center}
  \caption{Light curves, variations of flux ratio and flux--color
 diagrams in PKS~1510$-$089. Panel~(a) shows the light curves of the
 UV-bump subtracted data, the short-term and long-term components. 
 The flux is multiplied by a factor 5 for readability.
 Panel~(b) shows the $\nu\;F_{\nu\;V}/\nu\;F_{\nu\;J}$ of
 the UV-bump subtracted data, the short-term and the long-term
 components. We add 0.4 to the subtracted data of 
 $\nu\;F_{\nu\;V}/\nu\;F_{\nu\;J}$ for readability. Panel~(c), (d) and 
 (e) show the flux--color diagrams. In
 panel~(c), we show the diagram of the observed data. In panel~(d), we
 show the diagram of the UV-bump subtracted data. Panel~(e) is the
 diagram of the short-term component.}%
 \label{fig8} 
\end{figure}

As mentioned above, the color variations in both AO~0235$+$164 and
PKS~1510$-$089 can be explained by the two-component scenario; the
short-term flare component was superimposed on the long-term one having a
different color behavior. Such a two-component picture has been proposed
in past studies on blazar variability (e.g. \cite{Ghisellini97};
\cite{Villata04}; \cite{Sasada10}). The timescales of the long-term
components in AO~0235$+$164 and PKS~1510$-$089 are similar to those in
previous studies.

\subsection{Correlation between Amplitudes of Flux and Polarization
  Degree of Flares}
In section~3, we reported that there were positive correlations between
the flux and polarization degree in both AO~0235$+$164 and
PKS~1510$-$089. Such correlations between the flux and polarization
degree have been reported in other blazars in past studies
(e.g. \cite{Smith86}; \cite{Tosti98}), although the polarization
behaviors have generally been reported to be erratic
(e.g. \cite{Angel80}). We propose that short-term flares on a timescale
of 10~d in blazars are associated with increases in polarization degree,
with changes of the polarization angle of each flare. If a small flare
occurs, this flare is buried by other radiation component, for example
the long-term component as mentioned above, and then, the
polarization degree is hardly changed. However, a large amplitude flare
may not be buried like the flares ``A'', ``B'' and ``F''. In this
situation, the observed polarization degree should increase during the
large flare. In fact, the observed polarization degree during the large
flares ``A'', ``B'' and ``F'' have been changed dramatically.

We investigated the correlation between amplitudes of the flux and
polarization degree during flares. We used the photopolarimetric data of
41 blazars which we obtained from 2008 to 2010 \citep{Ikejiri09}. A full
description of the observation and the data reduction will be published
in a forthcoming paper. In this paper, we defined the flares and their
amplitudes of the flux and polarization degree as described
below. First, we defined the peak of a flare as the observation point
with the highest flux within $\pm$10~d. Second, we can calculated the
peak-to-valley amplitudes of the flux and polarization degree in this
20~d period. 

\begin{figure}
  \begin{center}
 \FigureFile(80mm,100mm){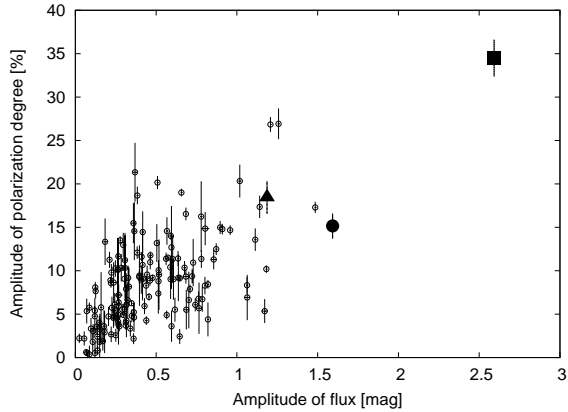}
  \end{center}
  \caption{Correlation between the amplitudes of the flux and
 polarization degree in short-term flares in blazars in the $V$
 band. Definitions of the amplitudes are described in the text. The
 filled circle, triangle and square represent the flares ``A'', ``B''
 and ``F'', respectively.}%
 \label{fig9} 
\end{figure}

Figure~9 shows correlation between the amplitudes of the flux and
polarization degree of short-term flares selected with the above
definition in the $V$ band. There is a positive correlation, and the
correlation coefficient is 0.66$^{+0.08}_{-0.10}$. The filled circle,
triangle and square represent the flares ``A'', ``B'' and ``F'' of
AO~0235$+$164 and PKS~1510$-$089, respectively. Those flares are one of
the largest flux- and polarization-amplitude flares in our sample
(figure~9). On the other hand, the correlation is weak in small flares;
there are flares having a relatively large amplitude of flux, but a
small amplitude of the polarization degree, and vice versa. This is
naturally expected since polarization variations of small flares are
readily buried by another variation component in our picture.

\citet{Sasada08} reported that a microvariation of S5~0716$+$714 had a
specific polarization component, and suggested that the microvariation
originated from a small and local region where the magnetic field was
aligned. Our result indicates that the short-term flares on timescales
of 10~d also originate from regions where the magnetic field is
aligned.

A part of previous observations of AO~0235$+$164 is apparently
inconsistent with the flux--polarization correlation that we obtained,
as mentioned in subsection~3.1. It may suggest that the correlation is
not a universal one. Details of variations are, however, unclear for
those previous observations; \citet{Takalo92} reported a quite low
polarization degree when the object was bright, while no other
observation is available within 40~d of it. \citet{Impey82} reported a
quite high polarization degree when the object was relatively faint,
while they reported only two observations. As mentioned in section~1,
the observed polarization is possibly a composition of multiple
polarization components having different timescales. The presence of
short- and long-term components is also indicated by the color
behavior in AO~0235$+$164, as discussed in the last subsection. The
exceptions deviating from the correlation between the flux and the
polarization degree in figure~2  might be due to the strong contribution
of a polarization component having a variation timescale longer than
that of the short flares. We need to make continuous photometric and
polarimetric monitoring of blazars with higher sampling rate for
understanding their polarization behavior more deeply.

\section{Conclusion}
We performed photometric and polarimetric monitoring of the blazars
AO~0235$+$164 and PKS~1510$-$089, and observed their flares. The large
flares ``A'', ``B'' and ``F'' exhibited the bluer-when-brighter
trends. The $V-J$ color variation of PKS~1510$-$089 showed the
redder-when-brighter trend, except for the period of the flare
``F''. The flux and color variations suggest that there were two
variation components having different timescales and colors. There were
positive correlations between the flux and polarization degree in both
objects. We studied the correlation between the amplitudes of the flux
and polarization degree of short flares in 41 blazars. We found that
there was a significant positive correlation between them. We propose
that the short-term flares on timescales of 10~d originate from a region
where the magnetic field is aligned. Observed polarization variations
are occasionally erratic probably because polarization variations of
small flares are readily buried by another variations having different
polarization angles and/or timescales.
\\
\\
This work was partly supported by a Grand-in-Aid from the Ministry of
Education, Culture, Sports, Science, and Technology of Japan
(22540252,). M.~Sasada and M.~Yamanaka have been supported by the JSPS 
Research Fellowship for Young Scientists.


\begin{thebibliography}{}
\bibitem[{Abdo} {et~al.,} (2010a)]{Abdo10a}
  Abdo,~A.~A.,\ \etal\ 2010a, \nat, 463, 919
\bibitem[{Abdo} {et~al.,} (2010b)]{Abdo10b}
  Abdo,~A.~A.,\ \etal\ 2010b, \apj, 721, 1425
\bibitem[{Angel} {\&} {Stockman} (1980)]{Angel80}
  Angel,~J.~R.~P.\ \& Stockman,~H.~S.\ 1980, \aap, 18, 321
\bibitem[{Burbidge} {et~al.,}(1976)]{Burbidge76}
  Burbidge, E. M., Caldwell, R. D., Smith, H. E., Liebert, J. \&
				    Spinrad, H. 1976, \apj, 205, L117
\bibitem[{Dermer} {\&} {Schlickeiser}(1993)]{Dermer93}
  Dermer,~C.~D.\ \& Schlickeiser,~R.\ 1993, \apj, 416, 458
\bibitem[{Fossati} {et~al.,}(1998)]{Fossati98}
  Fossati,~G.,\ Maraschi,~L.,\ Celloti,~A., Comastri,~A. \& Ghisellini,~G.\ 
                                    1998, \mnras, 299, 433
\bibitem[{Fukugita,} {Simasaku} {\&} {Ichikawa} (1995)]{Fukugita95}
  Fukugita,~M.,\ Shimasaku,~K.\ \& Ichikawa,~T.\ 1995, \pasp, 107, 945
\bibitem[{Ghisellini} {et~al.,}(1997)]{Ghisellini97}
  Ghisellini, G.,\ \etal\ 1997, \aap, 327, 61
\bibitem[{${\rm Gonz\acute{a}lez-P\acute{e}rez}$}
                           {et~al.,}(2001)]{Gonzalez-perez01}
                           ${\rm Gonz\acute{a}lez-P\acute{e}rez}$,~J.~N.,\
                           Kidger,~M.~R.\ \&
                           ${\rm Mart\acute{\i}n-Luis}$,~F.\
                           2001, \aj, 122, 2055
\bibitem[{Ikejiri} {et~al.,}(2009)]{Ikejiri09}
  Ikejiri,~Y.,\ \etal\ 2009, astro-ph/0912.3664
\bibitem[{Impey,} {Brand} {\&} {Tapia}(1982)]{Impey82}
  Impey, C. D.,\ Brand, P. W. J. L. \& Tapia, S. 1982, \mnras, 198, 1
\bibitem[{Jones,} {O'dell} {\&} {Stein}(1974)]{Jones74}
  Jones,~T.~W.,\ O'dell,~S.~L.\ \& Stein,~W.~A. 1974, \apj, 188, 353
\bibitem[{Jorstad} {et~al.,}(2007)]{Jorstad07}
  Jorstad,~S.~G.,\  \etal\ 2007, \aa, 134, 799
\bibitem[{Kataoka} {et~al.,}(2008)]{Kataoka08}
  Kataoka,~J.,\ \etal\ 2008, \apj, 672, 787
\bibitem[{Marscher} {\&} {Gear}(1985)]{Marscher85}
  Marscher,~A.~P.\ \& Gear,~W.~K.\ 1985, \apj, 298, 114
\bibitem[{Marscher} {et~al.,}(2008)]{Marscher08}
  Marscher,~A.~P.,\ \etal\ 2008, \nat, 452, 966
\bibitem[{Marscher} {et~al.,}(2010)]{Marscher10}
  Marscher,~A.~P.,\ \etal\ 2010, \apj, 710, L126
\bibitem[{Mead} {et~al.,}(1990)]{Mead90}
  Mead, A. R. G., Ballard, K. R., Brand, P. W. J. L., Hough, J. H.,
				      Brindle, C. \& Bailey, J. A. 1990,
				      \aap, 83, 183
\bibitem[{Miller} {et~al.,}(1981)]{Miller81}
  Miller,~H.~R. 1981, \apj, 244, 426
\bibitem[{Moore} {et~al.,}(1982)]{Moore82}
  Moore, R. L., Angel, J. R. P., Duerr, R., Lebofsky, M. J., Wisniewski, 
		W. Z., Rieke, G. H., Axon, D. J., Bailey, J., Hough, J. 
		M. \& McGraw, J. T. 1982, \apj, 260, 415
\bibitem[{Neugebauer,} {et~al.}(1979)]{Neugebauer79}
  Neugebauer,~G.,\ Oke,~J.~B.,\ Becklin,~E.~E.\ \& Matthews,~K. 1979, \apj, 230, 79
\bibitem[{Qian} {et~al.,}(1991)]{Qian91}
  Qian,~S.~J.,\ Quirrenbach,~A.,\ Witzel,~A.,\ Krichbaum,~T.~P.,\ Hummel,~
			   C.~A.\ \& Zensus,~J.~A.\ 1991, \aap,
			   241, 15 
\bibitem[{Raiteri} {et~al.,}(2001)]{Raiteri01}
  Raiteri, C. M.,\ \etal\ 2001, \aap, 377, 396
\bibitem[{Raiteri} {et~al.,}(2005)]{Raiteri05}
  Raiteri, C. M.,\ \etal\ 2005, \aap, 438, 39
\bibitem[{Raiteri} {et~al.,}(2006)]{Raiteri06}
  Raiteri, C. M.,\ \etal\ 2006, \aap, 459, 731
\bibitem[{Raiteri} {et~al.,}(2008)]{Raiteri08}
  Raiteri, C. M.,\ \etal\ 2008, \aap, 480, 339
\bibitem[{Rieke} {et~al.,}(1976)]{Rieke76}
  Rieke, G. H., Grasdalen, G. L., Kinman, T. D., Hintzen, P., Wills,
				    B. J. \& Wills, D. 1976, \nat, 260, 754
\bibitem[{Roberts} {et~al.,}(1976)]{Roberts76}
  Roberts,~M.~S., Brown,~R.~L., Brundage,~W.~D., Rots,~A.~H.,
		Haynes,~M.~P. \& Wolfe,~A.~M. 1976, \aj, 81, 293
\bibitem[{Sasada} {et~al.,}(2008)]{Sasada08}
  Sasada, M.,\ \etal\ 2008, \pasj, 60, L37
\bibitem[{Sasada} {et~al.,}(2010)]{Sasada10}
  Sasada, M.,\ \etal\ 2010, \pasj, 62, 645
\bibitem[{Sikora,} {Begelman} {\&} {Rees}(1994)]{Sikora94}
  Sikora,~M.,\ Begelman,~M.~C. \& Rees,~M.~J. 1994, \apj, 421, 153
\bibitem[{${\rm Sillanp\ddot{a}\ddot{a}}$} {et~al.,}(1993)]{Sillanpaa93}
  ${\rm Sillanp\ddot{a}\ddot{a}}$,~A.,\ Takalo,~L.~O.,\ Nilsson,~K.\
			   \& Kikuchi,~S.\ 1993, \apss, 206, 55 
\bibitem[{Singh,} {Shrader} {\&} {George}(1997)]{Singh97}
  Singh, K. P., Shrader,~C.~R. \& George,~I.~M. 1997, \apj, 491, 515
\bibitem[{Skrutskie} {et~al.,}(2006)]{Skrutskie06}
  Skrutskie,~M.~F.,\ \etal\ 2006, \aj, 131, 1163
\bibitem[{Smith,} {et~al.,} (1986)]{Smith86}
  Smith, P. S., Balonek, T. J., Heckert, P. A. \& Elston, R. 1986, \apj, 
		305, 484
\bibitem[{Takalo} {et~al.,}(1992)]{Takalo92}
  Takalo, L. O., ${\rm Sillanp\ddot{a}\ddot{a}}$, A., Nilsson, K., Kidger, M.,
				      de Diego, J. A. \& Piirola,
				      V. 1992, \aap, 94, 37
\bibitem[{Tosti} {et~al.,}(1998)]{Tosti98}
  Tosti, G.,\ \etal\ 1998, \aap, 339, 41
\bibitem[{Uemura} {et~al.,}(2010)]{Uemura10}
  Uemura,~M., \etal\ 2010, \pasj, 62, 69
\bibitem[{Villata} {et~al.,}(2004)]{Villata04}
  Villata, M.,\ \etal\ 2004, \aap, 421, 103
\bibitem[{Villforth} {et~al.,}(2010)]{Villforth10}
  Villforth,~C.,\ \etal\ 2010, \mnras, 402, 208
\bibitem[{Watanabe} {et~al.,}(2005)]{Watanabe05}
  Watanabe,~M.,\ \etal\ 2005, \pasp, 117, 870
\bibitem[{Wolff,} {Nordsieck} {\&} {Nook}(1996)]{Wolff96}
  Wolff,~M.~J.,\ Nordsieck,~K.~H. \& Nook,M.~A. 1996, \aj, 111, 856

\end{thebibliography}
\end{document}